\documentstyle[aps,twocolumn,epsf,prb,floats]{revtex}
\def\geqap{\,\raise 2pt \hbox{$>\kern-11pt \lower 5pt \hbox{$\sim$}$}\,}
\def\leqap{\,\raise 2pt \hbox{$<\kern-10pt \lower 5pt \hbox{$\sim$}$}\,}
\begin{document}
\draft
\twocolumn[\hsize\textwidth\columnwidth\hsize\csname @twocolumnfalse\endcsname
\title{Theory of Raman Scattering from Orbital Excitations in Manganese Oxides}
\author{S.~Okamoto$^1$, S.~Ishihara$^2$, and S.~Maekawa$^3$}
\address{$^1$The Institute of Physical and Chemical Research (RIKEN), Saitama 351-0198, Japan}
\address{$^2$Department of Applied Physics, University of Tokyo, Tokyo 113-8656, Japan}
\address{$^3$Institute for Materials Research, Tohoku University, Sendai 980-8577, Japan}
\date{\today}
\maketitle
\begin{abstract} 
We present a theory of the Raman scattering from the orbital wave excitations in manganese oxides. 
Two excitation processes of the Raman scattering are proposed. 
The Raman scattering cross section is formulated by using the pseudospin operator 
for orbital degree of freedom in a Mn ion. 
The Raman spectra from the orbital wave excitations are calculated and 
their implications in the recent experimental results reported in LaMnO$_3$ are discussed. 
\end{abstract}
\pacs{PACS numbers: 75.30.Et, 75.30.Vn, 71.10.-w, 78.20.Bh} 
]
\narrowtext
\noindent
\section{Introduction}
Since the discovery of the colossal magnetoresistance (CMR), 
much attention has been attracted to manganese oxides 
with perovskite structure.\cite{chahara,helmolt,tokura}
A variety of anomalous phenomena including gigantic decrease of the resistivity are 
observed in the vicinity of the phase transition from 
the charge and orbital ordered state to the forromagnetic metallic one in the oxides. 
One of the key factors to bring about the phenomena is 
the orbital degree of freedom in a Mn ion. \cite{science}
Due to the strong Hund coupling and crystalline field, 
two $e_g$ orbitals are degenerate and one of the $3d_{3z^2-r^2}$ and $3d_{x^2-y^2}$ orbitals 
is occupied by an electron in a Mn$^{3+}$ ion. 
\par
%
Extensive studies of the orbital ordering have been done. 
It is well known that orbital ordering associated with the Jahn-Teller (JT) type lattice distortion 
plays an important role to stabilize the layered (A) antiferromagnetic ordering in LaMnO$_3$, 
where spins align parallel (antiparallel) in the $xy$ plane (along the $z$ axis). 
\cite{goodenough,kanamori,matsumoto,hirota,ishihara1,murakami,rodriguez,maezono}
In the doped manganites, a variety of magnetic structures associated with 
the orbital orderings are reported. 
On the other hand, the dymanics of the orbital degree of freedom still remains to be clarified. 
In the orbital ordered state, the collective excitation of the orbital degree of freedom, 
termed orbital wave, has been theoretically predicted. 
\cite{ishihara1,cyrot,khaliullin,feiner,khaliullin2,brink,perebeinos}
The orbital wave corresponds to the collective electron excitation 
from occupied orbital to unoccupid one, 
i. e., the modulation of the shape of electronic cloud. 
When the orbital wave is excited, various low-energy properties, 
such as spin wave dispersion\cite{brink}, phonon dispersion\cite{perebeinos}, 
transport \cite{ishihara2} and thermodynamic properties, will be largely affected. 
However, the orbital wave excitations have not been observed 
because the experimental technique was limited. 
To clarify the dynamics of the orbital degree of freedom and 
its effects on the physical properties in manganese oxides, 
it is indispensable to establish a method to observe the orbital wave. 
\par
%
As a probe to observe the orbital wave, 
Ishihara {\it et al.} proposed the resonant inelastic x-ray scattering by which 
dispersion relation of the orbital wave can be detected.\cite{ishiharaRIXS} 
One of the other possible method to detect the orbital excitations is 
the Raman scattering.\cite{inoue} 
Although only the excitation at $\vec k=0$ and the density of states (DOS) may be detected, 
the Raman scattering has the following advantages: 
its energy resolution is of the order of 1~cm$^{-1}$ which is higher than that of the x-ray scattering 
and the different modes of the excitations are distinguished by 
the polarization analyses of the incident and scattered photons. 
Recently, the Raman scattering experiment in the detwinned single crystal of LaMnO$_3$ 
was carried out by Saitoh {\it et al}.\cite{e.saitoh} 
They found peak structures around 150~meV. 
By the detailed analyses of the polarization dependence and the Raman-shift energy of the spectra, 
the newly found spectra can be attributed to neither the multi-phonon nor the magnon excitations. 
Therefore, a theory of the Raman scattering from the remaining degree of freedom, 
i. e., the orbital excitation, is required to be developed. 
\par
In this paper, we present a detailed theoretical framework of the Raman scattering 
from the orbital-wave excitations in orbital ordered manganites. 
We propose two excitation processes. 
One of them is analogous to that in the two-magnon Raman scattering in antiferromagnets. 
However, the intensities of the Raman spectra from one- and two-orbital wave excitations are 
of the same order of magnitudes in this process, in contrast to the magnon Raman scattering. 
In another scattering process, photon induces exchange of electrons between Mn $e_g$ and O $2p$ orbitals, 
and one-orbital wave excitation is brought about. 
We formulate the scattering cross section by using the pseudospin operator 
for the orbital degree of freedom 
in Mn ions. 
Raman spectra from the orbital wave excitations are calculated 
as functions of energy in several polarization configurations. 
It is shown that the recent Raman experiments in LaMnO$_3$ are 
well explained by the orbital wave excitations. 
\par
In Sec.~II, the model Hamiltonian is introduced and dispersion relation of the orbital wave is 
investigated. 
In Sec.~III and Sec.~IV, 
the excitation processes for the Raman scattering are proposed and the cross section are formulated, 
respectively. 
Numerical results are presented in Sec.~V, 
where comparison between theory and experiment is shown. 
The last section is devoted to the summary and discussion. 
\section{Orbital Wave} 
Let us set up the model Hamiltonian describing the electronic state of manganese oxides.  
We consider the tight-binding Hamiltonian in the cubic lattice consisting of Mn ions. 
At each site, two $e_g$ orbitals are introduced  
and $t_{2g}$ electrons are treated as a localized spin ($\vec S_{t_{2g}}$) with $S_{t_{2g}}=3/2$.
We introduce three kinds of Coulomb interaction, i. e., 
the intra- ($U$) and inter-($U'$) orbital Coulomb interactions and the exchange interaction ($I$), 
between $e_g$ electrons at the same site. 
The energy splitting between two $e_g$ orbitals 
due to the JT distortion of a MnO$_6$ octahedron is represented by $g_{JT}\sqrt{Q_x^2+Q_z^2}$, 
where $g_{JT}$ and $Q_l$ are the electron-lattice coupling constant 
and the normal mode of the lattice distortion, respectively. 
The Hund coupling ($J_H$) between $e_g$ and $t_{2g}$ spins 
and the antiferromagnetic superexchange (SE) interaction ($J_{AF}$) between 
nearest neighboring (NN) $t_{2g}$ spins are introduced. 
Among these energy parameters, 
the intra-site Coulomb interactions are the largest.  \cite{t.saitoh}
Thus, by excluding the doubly occupied states in the $e_g$ orbitals, 
the following effective Hamiltonian for the low energy electronic state is derived:
\cite{ishihara1} 
\begin{eqnarray}
\widetilde{\cal H}= {\cal H}_J + {\cal H}_H + {\cal H}_{AF} + {\cal H}_{JT}.
\label{eq:heff}
\end{eqnarray}
The first term represents the SE interaction between NN $e_g$ electrons given by  
\begin{eqnarray}
\lefteqn{{\cal H}_J} \nonumber \\
&=& -2J_1\sum_{\langle ij \rangle } 
 \biggl ( {3 \over 4} n_i n_j + \vec S_i \cdot \vec S_j \biggr )
 \biggl ( {1 \over 4}  - \tau_i^l \tau_j^l \biggr ) \nonumber \\
&&-2J_2\sum_{\langle ij \rangle } 
 \biggl ( {1 \over 4} n_i n_j  - \vec S_i \cdot \vec S_j   \biggr )
 \biggl ( {3 \over 4}   + \tau_i^l \tau_j^l +\tau_i^l+\tau_j^l \biggr ), 
\label{eq:hj}
\end{eqnarray}
where $J_1$ and $J_2$ are the parameters for the SE interaction. 
Their definitions are given in Ref.~\onlinecite{JJ}. 
$\vec{S}_i$ is the spin operator of the $e_g$ electron with $S=1/2$. 
$\tau_i^l = \cos (\frac{2\pi}{3}m_l) T_{iz} - \sin (\frac{2\pi}{3}m_l) T_{ix}$ 
with $(m_x, m_y, m_z) = (1, -1, 0)$. 
$l$ denotes a direction of a bond between site $i$ and site $j$. 
$\vec{T}_i$ is the pseudospin operator for the orbital degree of freedom defined by 
$\vec{T}_i = (1/2) \sum_{\gamma \gamma' \sigma} 
\tilde{d}_{i \gamma \sigma}^{\dagger}(\vec{\sigma})_{\gamma \gamma'}\tilde{d}_{i \gamma' \sigma}$ 
where 
$\widetilde{d}_{i \gamma \sigma}$ is the annihilation operator of the $e_g$ electron 
at site $i$ with spin $\sigma$ and orbital $\gamma$. 
This operator excludes the doubly occupied states of electrons. 
$\langle T_{iz} \rangle= +(-)1/2$ corresponds to the state where the $d_{3z^2-r^2}$ ($d_{x^2-y^2}$) 
orbital is occupied by an electron.  
The second and third terms in Eq.~(\ref{eq:heff}) are given by 
\begin{eqnarray}
{\cal H}_{H} + {\cal H}_{AF}
= -J_H \sum_i \vec{S}_{i} \cdot \vec{S}_{t_{2g} i} 
+J_{AF} \sum_{\langle i j \rangle} \vec{S}_{t_{2g} i} \cdot \vec{S}_{t_{2g} j} . 
\label{eq:hhhaf}
\end{eqnarray}
Here, 
the anisotropy of the SE interaction originating from the tetragonal ($D_{4h}$) lattice distortion
is taken into account as 
$\sqrt{J_{1(2)}^z/J_{1(2)}^x}=\sqrt{J_{AF}^z/J_{AF}^x}=t_0^z/t_0^x=R$, 
$J_{1(2)}^x=J_{1(2)}^y$ and $J_{AF}^x=J_{AF}^y$. 
$t_0^{z(x)}$ is the transfer intensity between NN $d_{3z^2-r^2}(d_{3x^2-r^2})$ orbitals 
along the $z(x)$ axis. 
The last term in Eq.~(\ref{eq:heff}) is given by 
\begin{eqnarray}
{\cal H}_{JT}
= - g_{JT} \sum_{i \, l=z,x} Q_{i l} T_{i l} .
\label{eq:hjt}
\end{eqnarray}
$Q_{i \l}$ is defined by $(Q_{i z}, Q_{i x}) = Q (\cos \theta_i^{JT}, \sin \theta_i^{JT})$, 
where $\theta_i^{JT}$ represents 
the mixing of two normal modes of the lattice distortion. 
The lattice degree of freedom is assumed to be frozen in this paper, 
since the electronic process which we are interested in here has much 
larger energy than the lattice excitations. 
%
%
\par
\begin{figure}
\epsfxsize=0.7\columnwidth
\centerline{\epsffile{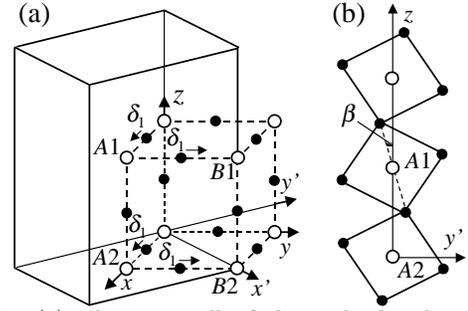}}
%
%
\caption{(a) The unit cell of the orthorhombic structure (straight lines). 
Open and filled circles represent Mn and O ions, respectively.  
The unit cell of the cubic perovskite structure is shown by broken lines.
Arrows indicate the displacements of O ions in the orthorhombic structures 
where $\delta_1$ is an amplitude of the displacement. 
(b) The alternate rotations of MnO$_6$ octahedra around $x'$ axis. 
$\beta$ is an angle of the rotation. 
}
\label{fig:fig1}
\end{figure}

We investigate the orbital states at the paramagnetic and $A$-AF phases 
by applying the mean field approximation. \cite{okamoto}
A unit cell which includes four Mn sites is adopted. 
These Mn sites are termed $A_1, A_2, B_1$ and $B_2$ (see Fig.~\ref{fig:fig1}). 
$A(B)$ and 1(2) classify the orbital and spin sublattices, respectively. 
As spin order parameters, we introduce 
$\langle S_{iz} \rangle$ and 
$\langle S_{t_{2g}iz} \rangle (=3\langle S_{iz} \rangle)$, 
where $\langle \cdots \rangle$ represents the thermal average. 
For the JT distortion of MnO$_6$ octahedra, 
$C$-type ordering with $(\theta_{A1(2)}^{JT}, \theta_{B1(2)}^{JT})=(2\pi/3, -2\pi/3)$ is adopted 
by considering the observed lattice distortion in LaMnO$_3$. 
For the orbital degree of freedom, 
we introduce the rotating frame and adopt the order parameter as 
$\langle \widetilde T_{iz} \rangle =
\cos \theta_i^t \langle T_{iz} \rangle + \sin \theta_i^t \langle T_{ix} \rangle$. 
$\theta_i^t$ describes the orbital state at site $i$ as 
$| \theta_i^t \rangle = 
\cos {\theta_i^t  \over 2} | 3z^2-r^2 \rangle + \sin {\theta_i^t  \over 2} | x^2-y^2 \rangle$. 
By minimizing the energy with respect to \{$\theta_i^t$\}, 
we obtain $(\theta_{A1(2)}^{t}, \theta_{B1(2)}^{t}) = (\theta_{A}, -\theta_{A})$ 
in both the paramagnetic and $A$-AF phases. 
%
%
%
%
%
\par
\begin{figure}
\epsfxsize=0.8\columnwidth
\centerline{\epsffile{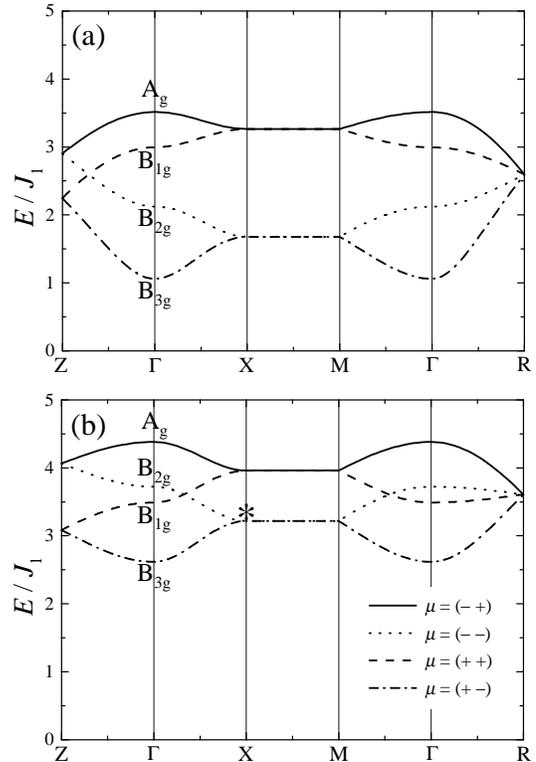}}
%
%
\caption{The dispersion relation of orbital wave in 
(a) paramagnetic and (b) $A$-AF phases. 
The Brillouin zone for the tetragonal lattice is adopted. 
Parameter values are chosen to be $J_2/J_1=0.35, R=1.07, g_{JT}Q/J_1=0.7$, 
and $\theta_{A1}^{JT}=2\pi/3$. 
Orbital state for the paramagnetic ($A$-AF) phase is denoted by 
$(\theta_{A1(2)}, \theta_{B1(2)})=(\theta_A, -\theta_A)$ 
with $\theta_A= 0.52 \pi (0.47 \pi)$.
An asterisk in (b) shows the mode which becomes Raman active 
when the monoclinic lattice distortion is introduced (See Fig.~\ref{fig:fig7}).}
\label{fig:fig2}
\end{figure}

The collective orbital excitations in the orbital ordered state are studied 
by applying the Holstein-Primakoff transformation to the pseudospin operators. 
Here, spins are assumed to be frozen. 
The orbital pseudospin is represented by using the bosonic operator, $a_i$ at site $i$, as 
$\tilde{T}_{iz}=1/2-a_i^{\dagger}a_i$ and $\tilde{T}_{ix}=1/2(a_i^{\dagger}+a_i)$. 
In Figs.~\ref{fig:fig2}~(a) and (b), 
the dispersion relation of the orbital waves in the paramagnetic and $A$-AF states are shown. 
Parameter values are chosen to be $J_2/J_1=0.35, R=1.07$ and $g_{JT}Q/J_1=0.7$. 
$J_1$ is estimated to be about 50~meV from the dispersion relation of the spin wave, 
the N{\'e}el temperature for $A$-AF and the orbital ordering temperature. 
It is shown that the orbital excitation has a gap in both cases. 
However, we note the origin of these gaps are different as follows: 
The gaps in paramagnetic and $A$-AF phases are given by 
$\sqrt{{\sqrt{3}\over2}g_{JT}Q({9\over2}J_1-{1\over2}J_2+{\sqrt{3}\over2}g_{JT}Q)}$ and
$\sqrt{(3J_1+{\sqrt{3}\over2}g_{JT}Q)(J_1+J_2+{\sqrt{3}\over2}g_{JT}Q)}$, respectively, 
where $\theta_{A}$ is chosen to be $\pi/2$. 
While the gap in the paramagnetic phase decreases with decreasing $g_{JT}Q$, 
the gap in the $A$-AF phase remains finite. 
This is because the latter originates from the anisotropic magnetic structure.\cite{ishihara1} 
$\mu=(\pm \pm')$ in Fig.~\ref{fig:fig2} denotes the mode of the orbital wave, 
where $\pm (\pm')$ represents the relative phase of 
the Holstein-Primakoff bosons of the four orbital sublattices in the $xy$ plane (along the $z$ axis).
Among four modes, $(-+)$ is the highest and its eigenoperator is given by 
\begin{eqnarray}
\alpha_{k}^{(-+)}=
\cosh \theta^{(-+)}(a_{A1\,k} - a_{B1\,k} + a_{A2\,k} - a_{B2\,k}) \nonumber \\ 
+\sinh \theta^{(-+)}(a_{A1\,-k}^{\dagger} - a_{B1\,-k}^{\dagger}
+ a_{A2\,-k}^{\dagger} - a_{B2\,-k}^{\dagger}) .
\label{eq:4modes}
\end{eqnarray}
$\cosh \theta^{(-+)}$ and $\sinh \theta^{(-+)}$ are the coefficients of the Bogoliubov transformation. 
This eigenoperator includes the following linear combinations of pseudospin operators: 
\begin{eqnarray}
&&\!\!\!\!\! \alpha_{k}^{(-+)} \nonumber \\
= && \sqrt{N_m \over N}\sum_{l(unit\, cell)} \!\!\! e^{i\vec k \vec r_l}
\Bigl( 
\cos \theta_A (\cosh \theta^{(-+)}+\sinh \theta^{(-+)}) \nonumber \\
&& \times (T_{A1(l)\,x} + e^{i k_y} T_{B1(l)\,x} 
+ e^{-i k_z} T_{A2(l)\,x} \nonumber \\ 
&& \hspace{13em} +e^{-i(k_y +k_z)} T_{B2(l)\,x}) \nonumber \\
&& -i(\cosh \theta^{(-+)}-\sinh \theta^{(-+)}) \nonumber \\
&& \times (T_{A1(l)\,y} - e^{i k_y}T_{B1(l)\,y} 
+ e^{-i k_z} T_{A2(l)\,y} \nonumber \\
&& \hspace{12em} - e^{-i(k_y +k_z)} T_{B2(l)\,y}) 
\Bigr) . 
\label{eq:alphax}
\end{eqnarray}
$N_m$ is the number of sites in the unit cell and the lattice constants are taken to be unity. 
By using the Group theory, it is shown that 
the irreducible representations of the modes $(-+)$, $(+-)$, $(--)$ and $(++)$ 
are identified as $A_g$, $B_{1g}, B_{3g}$ and $B_{2g}$ in the $D_{2h}$ group, respectively. 
From the irreducible representations, 
the allowed-mode symmetries of the orbital waves in several polarization configurations
are assigned. 
Throughout this paper, the polarization configuration is denoted by $(\zeta,\eta)$ 
where $\zeta$ and $\eta$ ($=x,y,z,x',y'$) represent the polarization of 
the incident and scattered photons, respectively. 
Here, $x,y$ and $z$ axes are the directions of the bonds connecting NN Mn sites 
and $x'\,(y')$ is defined by $x'=x+y\,(y'=-x+y)$. 
$x',z$ and $-y'$ axes correspond to $a,b$ and $c$ axes in $Pnma$ structure, respectively. 
(see Fig.~\ref{fig:fig1}) 
Polarization configurations and the allowed-mode symmetries of the orbital waves 
are obtained as follows: 
\begin{eqnarray}
(x,x) &\rightarrow& A_g+B_{2g}, \nonumber \\
(y,y) &\rightarrow& A_g+B_{2g}, \nonumber \\
(z,z) &\rightarrow& A_g, \nonumber \\
(x,y) &\rightarrow& A_g, \nonumber \\
(z,x) &\rightarrow& B_{2g}+B_{3g}, \nonumber \\
(x',x') &\rightarrow& A_g, \nonumber \\
(x',y') &\rightarrow& B_{2g}. \nonumber
\end{eqnarray}
%
%
%
%
%
%
\section{Scattering processes} 
In this section, we propose the excitation processes of the orbital wave in the Raman scattering. 
Here, we consider the Raman scattering experiment carried out by using the visible light. 
It has been reported that the electronic energy band in the region of 0$\sim$3~eV below the Fermi level 
in LaMnO$_3$ consists of Mn $e_g$ and O $2p$ orbitals.\cite{t.saitoh}
Therefore, it is expected that 
both the Mn $e_g$ and O $2p$ orbitals play important roles in the excitation processes. 
Taking into account these facts, the following two excitation processes are considered. 
\par
\begin{figure}
\epsfxsize=0.7\columnwidth
\centerline{\epsffile{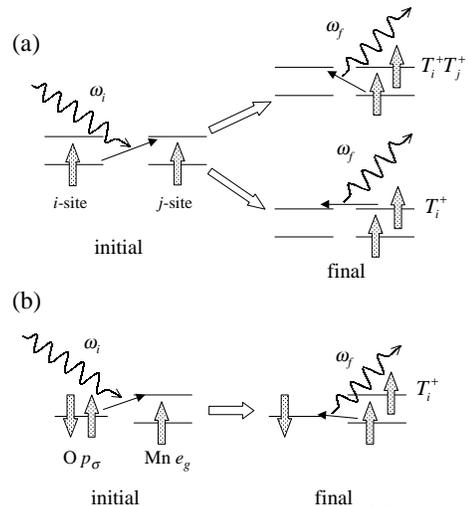}}
%
%
\caption{The excitation processes in the (a) $d$-$d$ and (b) $d$-$p$ processes.
Wavy lines represent the incident and scattered photons with 
energy $\omega_i$ and $\omega_f$, respectively. 
$i$ and $j$ in (a) represent NN Mn sites. 
O $p_{\sigma}$ in (b) represents one of the 6 oxygen $p_{\sigma}$ orbitals surrounding a Mn ion.
}
\label{fig:fig3}
\end{figure}

One of the processes is analogous to that in the two-magnon Raman scattering in the antiferromagnet. 
Mn $e_g$ orbitals are considered and O $2p$ orbitals are integrated out. 
The schematic picture of this process is presented in Fig.~\ref{fig:fig3}~(a). 
An electron at a Mn $e_g$ orbital is excited to one of 
NN $e_g$ orbitals through the interaction with an incident photon whose energy, momentum and 
polarization are denoted as $\hbar \omega_i$, $\vec k_i$ and $\lambda_i$, respectively. 
By emitting a photon with $\hbar \omega_f$, $\vec k_f$ and $\lambda_f$, 
one of the two electrons at a doubly-occupied Mn site returns to the empty Mn site. 
When the orbital states at one of the two or both the sites in the final state are different from 
those in the initial state, one- or two-orbital wave excitations are brought about, respectively. 
Hereafter, this process is termed the $d$-$d$ process. 
It is stressed that the scattering intensities 
from one- and two-orbital wave excitations are of the same order of magnitudes. 
This is due to the transfer intensity between the different orbitals at NN Mn sites. 
This characteristic is highly in contrast to the magnon Raman scattering 
in the antiferromagnet. \cite{fleury,elliott,shastry}
\par
In addition to the $d$-$d$ process, an orbital excitation is brought about 
through the exchange of electrons between Mn $e_g$ and O $2p$ orbitals. 
A schematic picture of this process is presented in Fig.~\ref{fig:fig3}~(b).
An electron in an O $p_{\sigma}$ orbital is excited to the neighboring Mn $e_g$ orbital, 
where $p_{\sigma}$ represents the O $2p$ orbital mixing with the Mn $e_g$ one through the $\sigma$ bond. 
There are two electrons in this Mn site and one hole in an O $p_{\sigma}$ orbital 
in the intermediate state. 
Then, one of the two electrons in this Mn site returns to the O site by emitting a photon. 
When the occupied orbital in the Mn site in the final state is different from that in the initial state, 
one-orbital wave excitation is brought about. 
Hereafter, this process is termed the $d$-$p$ process. 
In contrast to either the $d$-$d$ process or the magnon Raman scattering in the antiferromagnet, 
one-orbital wave is excited in the $d$-$p$ process. 
%
%
%
%
%
\section{Raman cross section}
In this section, we present the formulation for the cross section in the $d$-$d$ and $d$-$p$ processes. 
\subsection{$d$-$d$ process}
The cross section for the orbital wave excitations is calculated in 
the system where the $e_g$ electrons and photons are coupled. 
We adopt the following model Hamiltonian: 
\begin{eqnarray}
{\cal H}_{d-d}={\cal H}_e^{d-d}+{\cal H}_t^{d-d}+{\cal H}_{ph}+{\cal H}_{e-ph}^{d-d}.
\label{eq:h_inter}
\end{eqnarray}
The first and the second terms represent the intra-site electron-electron interactions 
and the electronic transfer between NN sites defined as 
\begin{eqnarray}
{\cal H}_e^{d-d}
&=&{\cal H}_H+\sum_{i \gamma \sigma} \varepsilon_d d_{i \gamma \sigma}^{\dagger} d_{i \gamma \sigma} 
\nonumber \\
&&+U \sum_{i \gamma} n_{i \gamma \uparrow} n_{i \gamma \downarrow} 
 +U'\sum_{i \gamma \sigma \sigma'} n_{i \gamma \sigma} n_{i -\gamma \sigma'} \nonumber \\ 
&&+I \sum_{i \gamma \sigma \sigma'} d_{i \gamma \sigma}^{\dagger} d_{i -\gamma \sigma}^{\dagger} 
   d_{i \gamma \sigma'} d_{i -\gamma \sigma}
\label{eq:h_e}
\end{eqnarray}
and
\begin{eqnarray}
{\cal H}_{t}^{d-d} 
&=&\sum_{{\langle ij \rangle} \atop {\gamma \gamma' \sigma}} \left(
t_{ij}^{\gamma \gamma'} d_{i \gamma \sigma}^{\dagger} d_{j \gamma' \sigma}+{\rm H.c.}\right),
\label{eq:h_tinter}
\end{eqnarray}
respectively. 
$d_{i \gamma \sigma}$ is the annihilation operator of $e_g$ electron 
at site $i$ with spin $\sigma$ and orbital $\gamma$. 
Its energy level is given by $\varepsilon_d$. 
$t_{ij}^{\gamma \gamma'}$ is the transfer intensity 
between $\gamma$ orbital at site $i$ and $\gamma'$ orbital at site $j$ 
and is obtained by the second-order perturbation with respect to 
the electron transfer between Mn~$e_g$ and O~2p orbitals. 
The explicit form of $t_{ij}^{\gamma \gamma'}$ is given by the Slater-Koster formulas. \cite{slater}
The third term represents the photon system as 
\begin{eqnarray}
{\cal H}_{ph}
=\sum_{\vec k \lambda} 
\hbar \omega_{\vec k} \left( b_{\vec k \lambda}^{\dagger} b_{\vec k \lambda} +{1\over2} \right), 
\label{eq:h_ph}
\end{eqnarray}
where $b_{\vec k \lambda}$ is the annihilation operator of photon 
with energy $\hbar \omega_{\vec k}$, momentum $\vec k$ and polarization $\lambda$. 
Finally, the electron-photon interaction is described by the fourth term in Eq.~(\ref{eq:h_inter}) as 
\begin{eqnarray}
{\cal H}_{e-ph}^{d-d}
=-{e \over c}\sum_{\langle i j \rangle \gamma \gamma'} 
{\vec A}(\vec r) \cdot {\vec j}_{ij}^{\gamma \gamma'} .
\label{eq:h_epinter}
\end{eqnarray}
Here, ${\vec A}(\vec r)$ and ${\vec j}_{ij}^{\gamma \gamma'}$ 
are the vector potential at $\vec r=(\vec r_i+\vec r_j)/2$ and the current operator, respectively. 
These are given by 
\begin{eqnarray}
{\vec A}(\vec r)
={1\over\sqrt{V}}\sum_{\vec k \lambda} \sqrt{2 \pi \hbar c^2 \over \omega_{\vec k}} {\hat e}_\lambda
\left(b_{\vec k \lambda}^{\dagger}e^{-i\vec k \vec r} +b_{\vec k \lambda}e^{i\vec k \vec r} \right) ,
\label{eq:vecp}
\end{eqnarray}
and 
\begin{eqnarray}
{\vec j}_{ij}^{\gamma \gamma'}
={i\over \hbar}\vec r_{ij}t_{ij}^{\gamma \gamma'}
\sum_\sigma d_{i \gamma \sigma}^\dagger d_{j \gamma' \sigma}+{\rm H.c.}\,.
\label{eq:j_dd}
\end{eqnarray}
$\hat e_{\lambda}$ is the unit vector along the polarization $\lambda$ 
and $\vec r_{ij}=(\vec r_i-\vec r_j)$. 
\par
Being based on the Hamiltonian ${\cal H}_{d-d}$, the scattering matrix is calculated. 
The initial and final states of the scattering are described by $\widetilde {\cal H}$ 
and the intermediate state is described by ${\cal H}_e^{d-d}$. 
The scattering matrix in the $d$-$d$ process is obtained as, 
\begin{eqnarray}
\lefteqn{S_{fi}^{d-d}} \nonumber \\
&=&{4 \pi C i \over V} \delta(E_f-E_i)
\sum_{\langle ij \rangle} 
\biggl(\hat e_{\lambda f} \cdot {\vec r_{ij} \over l_{dd}} \biggr) 
\biggl(\hat e_{\lambda i} \cdot {\vec r_{ij} \over l_{dd}} \biggr) 
\nonumber \\
&&\!\!\!\times \Bigg( 2\tilde J_1
 \biggl ( {3 \over 4} n_i n_j + \vec S_i \cdot \vec S_j   \biggr )
 \biggl ( {1 \over 4}  - \tau_i^l \tau_j^l \biggr ) \nonumber \\
&&\:+2\tilde J_2 
 \biggl ( {1 \over 4} n_i n_j  - \vec S_i \cdot \vec S_j   \biggr )
 \biggl ( {3 \over 4}   + \tau_i^l \tau_j^l +\tau_i^l+\tau_j^l \biggr ) \Biggr) , 
\label{eq:sinter}
\end{eqnarray}
with  
$C= \pi e^2 / (\hbar^2 \sqrt{\omega_{\vec k_i} \omega_{\vec k_f}})$. 
$E_{i(f)}$ represents the energy of the initial (final) state.
${\tilde J}_1$ and ${\tilde J}_2$ are defined by 
${\tilde J}_1=l_{dd}^2 t_0^2/(U'-I+ \hbar \omega_{\vec k_f})
+l_{dd}^2 t_0^2/(U'-I- \hbar \omega_{\vec k_i})$ 
and 
${\tilde J}_2=l_{dd}^2 t_0^2/(U+2J_H+ \hbar \omega_{\vec k_f})
+l_{dd}^2 t_0^2/(U+2J_H- \hbar \omega_{\vec k_i})$, respectively, 
where $l_{dd}$ is the distance between NN Mn sites. 
Finally, by rewriting the pseudospin operators in Eq.~(\ref{eq:sinter}) by using 
the Holstein-Primakoff bosons, 
we obtain the cross section in the $d$-$d$ process as 
\begin{eqnarray}
\lefteqn{I_1^{d-d} = 
{\omega_f^2 C^2 N \over \hbar (2 \pi c^2)^2}
{4 \over N_m}} \nonumber \\
&& \times \sum_{\mu=1}^{N_m} \left| \sum_{\rho \nu \nu'}
(\hat e_{\lambda f}\cdot \vec \rho)(\hat e_{\lambda i}\cdot \vec \rho)
K_{\nu \nu'}^{\rho} (V_{\nu \mu}(0)+W_{\nu \mu}(0)) \right|^2 \nonumber \\
&& \times \left( n_{\mu 0}\delta(\mit\Delta E+\varepsilon_{\mu 0})
+(1+n_{\mu 0})\delta(\mit\Delta E-\varepsilon_{\mu 0}) \right). 
\label{eq:w1inter}
\end{eqnarray}
$n_{\mu \vec k}$ is the number of the boson of mode $\mu$, momentum $\vec k$ 
and energy $\varepsilon_{\mu \vec k}$.
$V_{\nu \mu} (\vec k)$ and $W_{\nu \mu} (\vec k)$ are the coefficients of the Bogoliubov transformation 
connecting the boson operator for the $\nu$ th ion to the $\mu$ th eigenmode as
\begin{eqnarray}
a_{\nu \vec k} = 
V_{\nu \mu} (\vec k) \alpha_{\mu \vec k} 
+W_{\nu \mu} (\vec k) \alpha_{\mu -\vec k}^{\dagger} \, .
\end{eqnarray}
$K_{\nu \nu'}^{\rho}$ is given by  
\begin{eqnarray}
K_{\nu \nu'}^{\rho} =
\biggl( 
\tilde J_1 \Bigl( {3 \over 4} + S_{\rho} \Bigr)
-\tilde J_2 \Bigl( {1 \over 4} - S_{\rho} \Bigr)
\biggr)
S_{\nu}^{\rho} C_{\nu'}^{\rho} \nonumber \\
-\tilde J_2 \Bigl( {1 \over 4} - S_{\rho} \Bigr) S_{\nu}^{\rho}, 
\label{eq:K}
\end{eqnarray}
with $S_{\rho}=\langle \vec S_i \cdot \vec S_{i+\vec \rho} \rangle$. 
$S_{\nu}^{\rho}$ and $C_{\nu}^{\rho}$ are defined by 
$S_{\nu}^{\rho}=\sin(\theta_{\nu}^t+{2 \pi \over3}m_{\rho} )$ and 
$C_{\nu}^{\rho}=\cos(\theta_{\nu}^t+{2 \pi \over 3}m_{\rho})$, respectively. 
$\mit\Delta E(=\hbar \omega_{\vec k_i}-\hbar \omega_{\vec k_f})$ is the Raman shift energy. 
The cross section for the two-orbital wave excitation is given by 
\begin{eqnarray}
\lefteqn{I_2^{d-d} = 
{\omega_f^2 C^2 N \over \hbar (2 \pi c^2)^2}
{4 \over N_m}}\nonumber \\ 
&&\times \sum_{\vec k \mu \mu'}
\Biggl| \sum_{\vec \rho  \nu \nu'}
(\hat e_{\lambda f}\cdot \vec \rho)(\hat e_{\lambda i}\cdot \vec \rho)
\Bigl[
L_{\nu \nu'}^{\rho} V_{\nu \mu}(\vec k) W_{\nu \mu}(-\vec k) \nonumber \\
&&\hspace{4em}+
M_{\nu \nu'}^{\rho}(\vec k) \delta_{\nu, \nu'+\rho} 
( V_{\nu \mu}(\vec k) + W_{\nu \mu}(\vec k) ) \nonumber \\
&&\hspace{8em}\times ( V_{\nu' \mu'}(-\vec k) + W_{\nu' \mu'}(-\vec k) ) 
\Bigr]
\Biggr|^2 \nonumber \\
&&\quad \times  (1+n_{\mu \vec k})(1+n_{\mu' -\vec k}) 
\delta(\mit\Delta E-\varepsilon_{\mu \vec k}-\varepsilon_{\mu' -\vec k}) , 
\label{eq:w2inter}
\end{eqnarray}
with 
\begin{eqnarray}
L_{\nu \nu'}^{\rho} =
\biggl( 
\tilde J_1 \Bigl( {3 \over 4} + S_{\rho} \Bigr)
-\tilde J_2 \Bigl( {1 \over 4} - S_{\rho} \Bigr)
\biggr)
C_{\nu}^{\rho} C_{\nu'}^{\rho} \nonumber \\
-2 \tilde J_2 \Bigl( {1 \over 4} - S_{\rho} \Bigr) C_{\nu}^{\rho}, 
\label{eq:L}
\end{eqnarray}
and
\begin{eqnarray}
M_{\nu \nu'}^{\rho}(\vec k) =
2\biggl( 
\tilde J_1 \Bigl( {3 \over 4} + S_{\rho} \Bigr)
-\tilde J_2 \Bigl( {1 \over 4} - S_{\rho} \Bigr)
\biggr) \nonumber \\
\times S_{\nu}^{\rho} S_{\nu'}^{\rho} \cos k_{\rho}. 
\label{eq:M}
\end{eqnarray}
In Eq.~(\ref{eq:w2inter}), the anti-Stokes parts are neglected for simplicity. 
The interaction between two orbital waves in the final state is neglected. 
This interaction is expected to shift the two-orbital wave Raman spectrum to the lower energy region 
as the magnon-magnon interaction does in the two-magnon Raman scattering. \cite{elliott}
\subsection{$d$-$p$ process}
The cross section from the orbital wave excitation is calculated by using the system where 
Mn $e_g$ and O $p_{\sigma}$ orbitals and the electron-photon coupling are taken into account. 
We start with the following model Hamiltonian: 
\begin{eqnarray}
{\cal H}_{d-p}={\cal H}_e^{d-p}+{\cal H}_t^{d-p}+{\cal H}_{ph}+{\cal H}_{e-ph}^{d-p}. 
\label{eq:h_intra}
\end{eqnarray}
The first and second terms describe 
the intra-site electron-electron interactions and electron hopping as 
${\cal H}_e^{d-p}={\cal H}_e^{d-d}+
{1\over2}\sum_{i \delta \sigma} 
\varepsilon_{p} p_{i \delta \sigma}^{\dagger} p_{i \delta \sigma}$ and 
\begin{eqnarray}
{\cal H}_{t}^{d-p}
=\sum_{i \gamma \delta \sigma}\left( t_{\gamma \delta}
d_{i\gamma\sigma}^{\dagger} p_{i\delta\sigma}+{\rm H.c.}\right), 
\label{eq:h_tintra}
\end{eqnarray}
respectively, where 
$\varepsilon_{p}$ is the energy level of O $p_{\sigma}$ orbital and 
$p_{i \delta \sigma}$ is the annihilation operator of the $p_{\sigma}$ electron 
at $\vec r_i+l_{pd}\vec \delta$. 
$l_{pd}$ is the distance between NN Mn and O sites. 
$\vec \delta$'s are the unit vectors along $x,y$ and $z$ directions. 
$t_{\gamma \delta}$ represents the transfer intensity between NN Mn $e_g$ and O $p_{\sigma}$ orbitals 
and is given by 
\begin{eqnarray}
t_{\gamma \delta}
=t_{pd}\left(\begin{array}{ccc}
-\frac12       & -\frac12        & 1 \\
\frac{\sqrt3}2 & -\frac{\sqrt3}2 & 0 \\
\end{array}
\right)_{\gamma \delta} , 
\label{eq:tpd}
\end{eqnarray}
for $\delta=x,y,z$ with $t_{\gamma \delta}=-t_{\gamma -\delta}$.
$t_{pd}$ is the transfer intensity between Mn $d_{3z^2-r^2}$ orbital at $\vec r_i$ 
and O $p_z$ orbital at $\vec r_i+l_{pd}\hat z$. 
The electron-photon interaction is represented by the fourth term, 
\begin{eqnarray}
{\cal H}_{e-ph}^{d-p}
=-{e \over c}\sum_{i \gamma \vec \delta} 
{\vec A}(\vec r_i) \cdot {\vec j}_{i \gamma \delta}. 
\label{eq:h_epintra}
\end{eqnarray}
${\vec j}_{i \gamma \delta}$ is the current operator representing the transition between 
Mn $\gamma$ orbital at $\vec r_i$ and O $p_{\sigma}$ orbital at $\vec r_i + l_{pd} \vec \delta$ 
given by 
\begin{eqnarray}
{\vec j}_{i \gamma \delta}
={i\over \hbar} l_{pd}{\vec \delta} t_{\gamma \delta}
\sum_\sigma d_{i \gamma \sigma}^\dagger p_{i \delta \sigma} + {\rm H.c.} \,. 
\label{eq:j_intra}
\end{eqnarray}
\par
The scattering matrix is obtained by the second order perturbation 
with respect to ${\cal H}_{e-ph}^{d-p}$ 
and is given by 
\begin{eqnarray}
S_{fi}^{d-p} = {4 \pi C i \over V}
\delta(E_f-E_i)
\sum_{i \vec \rho} 
(\hat e_{\lambda f} \cdot \vec \rho)(\hat e_{\lambda i} \cdot \vec \rho)
\, \tilde J \, \tau_i^\rho, 
\label{eq:sintra}
\end{eqnarray}
where 
$\vec \rho = \hat x, \hat y, \hat z$ 
and 
\begin{eqnarray}
{\tilde J} &=& l_{pd}^2 t_{pd}^2 \Biggl(
{{3 \over 2} \over U'-I-{3\over4}J_H-\Delta+ \hbar \omega_{\vec k_f}} \nonumber\\
&&+{{3 \over 2} \over U'-I-{3\over4}J_H-\Delta- \hbar \omega_{\vec k_i}} 
-{{1 \over 2} \over U+{5\over4}J_H-\Delta+ \hbar \omega_{\vec k_f}} \nonumber \\
&&\hspace{10em}-{{1 \over 2} \over U+{5\over4}J_H-\Delta- \hbar \omega_{\vec k_i}} \Biggr),
\end{eqnarray}
with $\Delta=\varepsilon_p - \varepsilon_d$. 
It is worth to mention that $S_{fi}^{d-p}$ includes the linear term of $\tau_i$ 
because the orbital excitation is brought about in a MnO$_6$ octahedron. 
Therefore, one-orbital wave excitation contributes to the Raman scattering. 
Finally, by rewriting the pseudospin operators in Eq.~(\ref{eq:sintra}) by 
the Holstein-Primakoff bosons, 
we obtain the cross section in the $d$-$p$ process as follows, 
\begin{eqnarray}
I^{d-p} &=& 
{\omega_f^2 C^2 N \over \hbar (2 \pi c^2)^2}
{\tilde J}^2 \nonumber \\
&\times& {4 \over N_m} \sum_{\mu}
\left|\sum_{\rho \nu}
(\hat e_{\lambda f}\cdot \vec \rho)(\hat e_{\lambda i}\cdot \vec \rho)
S_{\nu}^{\rho} (V_{\nu \mu 0}+W_{\nu \mu 0}) \right|^2 \nonumber \\
&\times&  \left( n_{\mu 0}\delta(\mit\Delta E-\varepsilon_{\mu 0})
+(1+n_{\mu 0})\delta(\mit\Delta E+\varepsilon_{\mu 0}) \right) . 
\label{eq:wintra}
\end{eqnarray}
\par
\section{Numerical Results}
\subsection{$d$-$d$ process}
%
%
\begin{figure}
\epsfxsize=1.0\columnwidth
\centerline{\epsffile{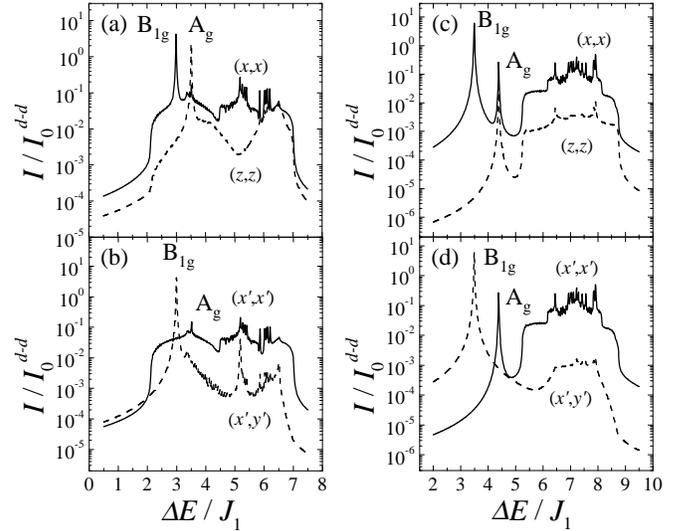}}
%
%
\caption{The Raman scattering spectra from the orbital waves with 
(a), (b) paramagnetic and 
(c), (d) $A$-AF spin structures in the $d$-$d$ process. 
$I_0^{d-d}$ is defined by $I_0^{d-d}=\omega_f^2 C^2 N / (\hbar (2\pi)^2) \tilde J_1^2$  and is assumed 
to be independent of $\omega_i$ and $\omega_f$, for simplicity.  
A value of $\tilde J_2 / \tilde J_1$ is chosen to be 0.35. 
Other parameter values are the same as those in Fig.~\ref{fig:fig2}. 
}
\label{fig:fig4}
\end{figure}
%
%
%
The Raman scattering spectra in the $d$-$d$ process are given 
by the summation of $I_1^{d-d}$ and $I_2^{d-d}$. 
Numerical results for the Raman spectra in the paramagnetic and $A$-AF phases 
are presented in Fig.~\ref{fig:fig4}. 
Sharp spectra in the region $3.0 < {\mit\Delta} E/J_1 < 3.6$ $(3.5 < {\mit\Delta} E/J_1 < 4.5)$ 
and broad ones in the region $2.0 < {\mit\Delta} E/J_1 < 7.0$ $(5.0 < {\mit\Delta} E/J_1 < 9.0)$
in the paramagnetic ($A$-AF) phase originate from one- and two-orbital wave excitations, respectively.
%
%
For the one-orbital wave excitations in the $(x,y)$ and $(z,x)$ polarizations, 
all modes are Raman inactive as shown in Fig.~\ref{fig:fig4}. 
In these configurations, 
polarization of the scattered photon is perpendicular to that of the incident photon and 
both the polarizations are parallel to the bonds between NN Mn sites. 
Therefore, in these configurations, matrix element $S_{fi}^{d-d}$ in Eq.~(\ref{eq:sinter}) 
vanishes and orbital excitations are prohibited. 
%
%
The polarization dependence of the Raman spectra from two-orbital wave excitation 
are obtained as follows: 
\begin{eqnarray}
\mbox{$(x,x),(z,z),(x',x')$ and $(x',y')$} \rightarrow \mbox{active}, \nonumber \\
\mbox{$(x,y)$ and $(z,x)$} \rightarrow \mbox{inactive}. 
\nonumber
\end{eqnarray}
The following relations of the relative intensity are shown: 
$I(x,x) \sim I(x',x') > I(z,z) \gg I(x',y')$. 
These relations reflect the type of orbital ordering, 
that is, the $C$-type with $\theta_{A1}^t = 0.507 \pi$ and $0.481 \pi$ 
for paramagnetic and $A$-AF phases, respectively. 
Due to this type of orbital ordering, SE interaction between NN $e_g$ electrons 
in the $xy$ plane is much stronger than that along the $z$ axis. 
Therefore, the intensity in the $(z,z)$ polarization is smaller than those 
in the $(x,x)$ and $(x',x')$ polarizations.
Small intensity in the $(x',y')$ configuration is 
attributed to the interference effect between the orbital waves with the different symmetry. 
In the spectra from two-orbital wave excitations, several peaks and edges are shown. 
As the two-magnon Raman spectra reflect the DOS of the magnon, 
the two-orbital wave Raman spectra reflects the DOS of the orbital waves; 
position of the each peak corresponds to the van~Hove singularity of DOS of the orbital waves. 
However, the spectra are not the DOS of the orbital waves itself because of the $k$ dependence of 
the matrix elements $M_{\nu \nu'}^{\rho}$. 
The $d$-$d$ process may be usefull to examine DOS of the orbital waves 
as well as its excitation with $\vec k=0$. 
\subsection{$d$-$p$ process}
\begin{figure}
\epsfxsize=1.0\columnwidth
\centerline{\epsffile{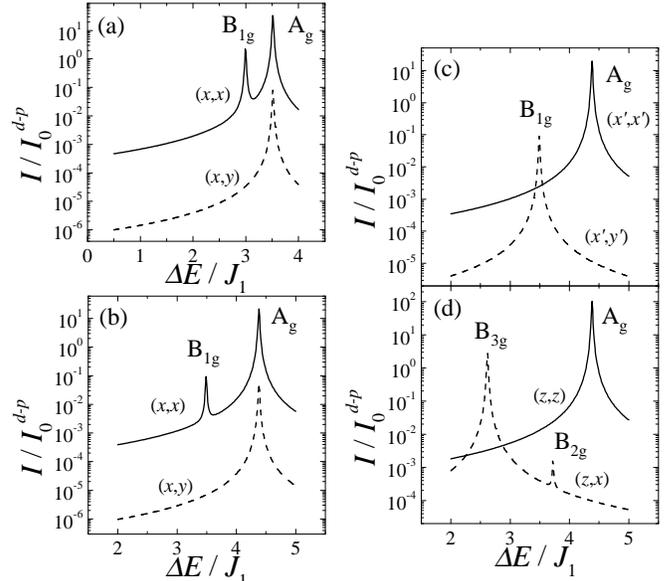}}
%
%
\caption{The Raman scattering spectra from the orbital waves in  
(a) the paramagnetic and (b)-(d) $A$-AF phases in the $d$-$p$ process. 
$I_0^{d-p}$ is defined by $I_0^{d-p}=\omega_f^2 C^2 N / (\hbar (2\pi)^2) \tilde J^2$ and is assumed 
to be independent of $\omega_{i,f}$, for simplicity. 
The displacement of the O ions and the rotation of MnO$_6$ octahedra are set to be 
$\delta_{1}/l_{pd}=0.04$ and $\beta={\pi \over 18}$. 
Other parameter values are the same as those in Fig.~\ref{fig:fig2}. 
}
\label{fig:fig5}
\end{figure}

In Fig.~\ref{fig:fig5}, numerical results of the Raman spectra for the $d$-$p$ process 
in the paramagnetic and $A$-AF phases are shown. 
We find that 
the relative intensity of the two spectra from the $A_g$ and $B_{1g}$ modes 
in the $(x,x)$ configuration in the $d$-$p$ process is different from that in the $d$-$d$ process, 
i. e., the spectrum of the $A_g$ mode becomes larger than that of the $B_{1g}$ mode 
in the $d$-$p$ process. 
As shown in Eq.~(\ref{eq:alphax}), in the $A_g$ mode, 
the orbital excitation at each site occurs in-phase. 
Therefore, interference effect in the in-phase $A_g$ mode increases its intensity 
in the $d$-$p$ process. 
On the other hand, in the $d$-$d$ process, orbital excitation is doinated by the process 
where the two electrons in the NN Mn sites are exchanged with each other. 
Because a minus sign in the scattering matrix arises from the exchange of electrons, 
interference effects do not occur. 
This reflects the AF-type orbital ordering in the ground state. 
%
%
%
%
%
\subsection{Effect of the lattice distortion}
In the previous subsections, 
the Raman scattering spectra have been calculated in the orthorhombic crystal structure. 
However, it is reported that the lattice structure of LaMnO$_3$ is monoclinic 
when the oxygen partial pressure during synthesis is reduced.\cite{mitchell} 
Actually, a sample used in the recent Raman scattering experiments shows 
the monoclinic structure.\cite{tsuda} 
Therefore, we examine the effect of the monoclinic distortion on the Raman spectra 
in order to compare the present theory with the experiments. 
\par
%
%
%
\begin{figure}
\epsfxsize=0.45\columnwidth
\centerline{\epsffile{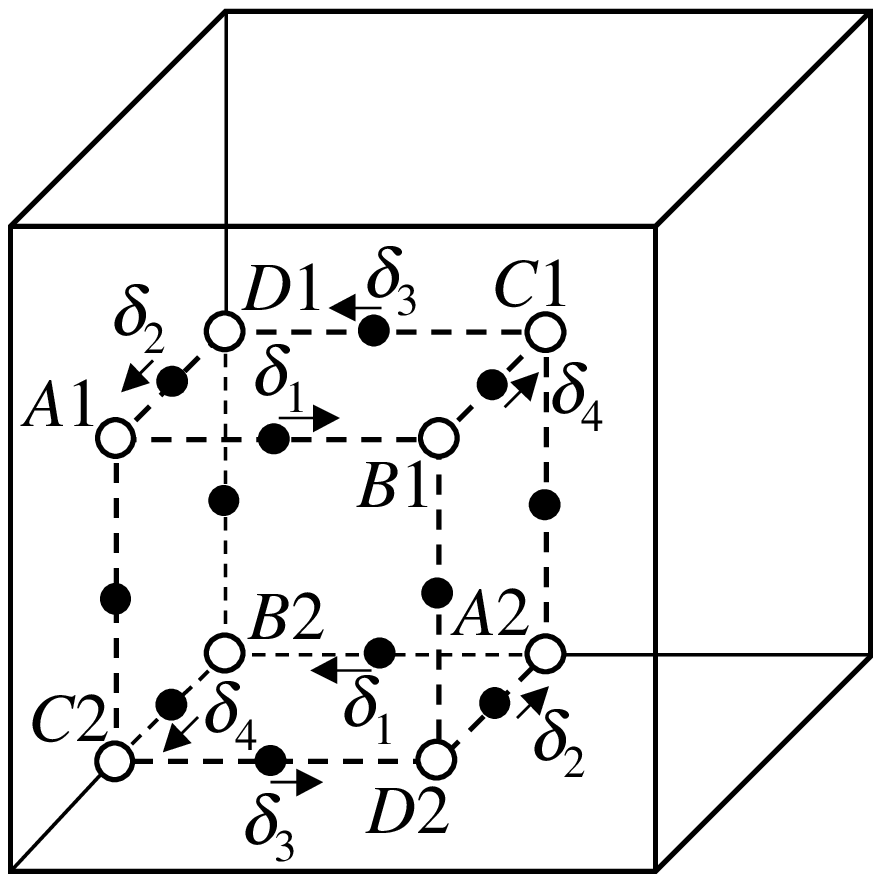}}
%
%
\caption{The unit cell of the monoclinic structures (straight lines). 
Open and filled circles represent Mn and O ions, respectively.  
The unit cell of the cubic perovskite structure is shown by broken lines.
The arrows indicate the displacement of O ions in the monoclinic structure 
where $\delta_{1 \sim 4}$ is the amplitude of the displacement. 
The alternate rotations of MnO$_6$ octahedra around $x'$ axis are shown in Fig.~\ref{fig:fig1}~(b). 
}
\label{fig:fig6}
\end{figure}
%
%
We adopt a model of the crystal structure which is schematically shown in Fig.~\ref{fig:fig6}. 
Displacements of the O ions indicated by arrows are taken into account 
up to the order of $O(\delta_i/l_{pd})$ as follows. 
The transfer intensity between NN Mn~$e_g$ orbitals in the $xy$ plane is given by 
$t_0^{xy}(\delta)=t_0(0)(1+\,\delta/l_{pd})^{3.5}(1-\,\delta/l_{pd})^{3.5}$. 
Therefore, $t_0^{xy}$ does not change within the order of $O(\delta_i/l_{pd})$. 
$J_{1,2}^{xy}$ does not change, either, because $J_{1,2}^{xy}$ are proportional to $(t_0^{xy})^2$. 
The transfer intensity between NN Mn~$e_g$ and O~$2p$ orbitals changes as 
$t_{pd}(\delta)=t_{pd}(0)(1 \pm 3.5\,\delta/l_{pd})$. 
From these facts, 
the dispersion relation of the orbital wave and the Raman spectra in the $d$-$d$ process 
are insensitive to the oxygen displacement
but the spectra in the $d$-$p$ process are sensitive. 
Therefore, in the following, we concentrate on the $d$-$p$ process. 
The dispersion relation of the orbital waves is calculated 
in the unit cell which includes 8 Mn sites. 
\par
\begin{figure}
\epsfxsize=0.7\columnwidth
\centerline{\epsffile{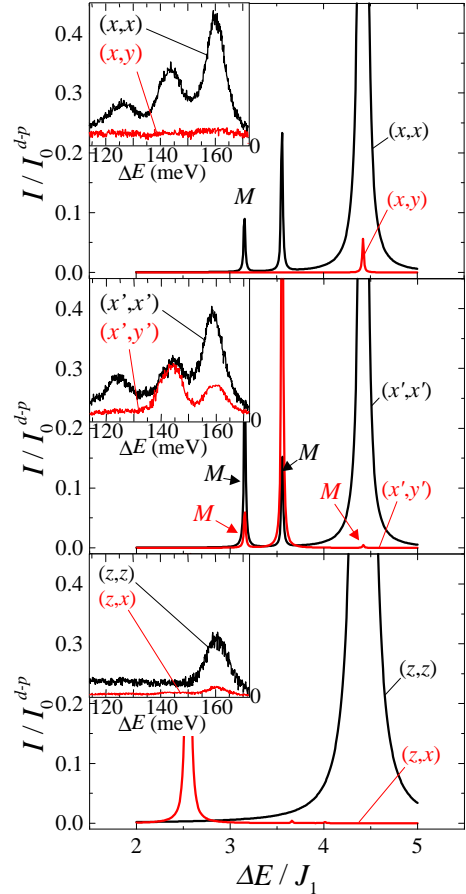}}
\caption{The Raman scattering spectra from the orbital wave in the $A$-AF phase. 
The $d$-$p$ process in the monoclinic lattice distortion is assumed.
The displacement of the O ions and the rotation of MnO$_6$ octahedra are chosen to be 
$\delta_{1}/l_{pd}=0.03, \delta_{2}/l_{pd}=0.11, \delta_{3}/l_{pd}=0.07, \delta_{4}/l_{pd}=0.15$
and $\beta={\pi \over 18}$. 
The anisotropy in the transfer intensity is chosen to be $R=1.15$. 
Other parameter values are the same as those in Fig.~\ref{fig:fig2}. 
Orbital state is given by 
$(\theta_{A1(2)}, \theta_{B1(2)}, \theta_{C1(2)}, \theta_{D1(2)}) 
=(\theta_A, -\theta_A, \theta_A, -\theta_A)$ 
with $\theta_A= 0.47 \pi$.
Insets show the experimental Raman spectra in LaMnO$_3$ at 9~K in Ref.~21. 
Vertical axes of the experimental data are arbitrary. 
}
\label{fig:fig7}
\end{figure}

In Fig.~\ref{fig:fig7}, the numerical results for the $d$-$p$ process in the monoclinic lattice 
are shown. 
Here, $A$-AF structure is adopted. 
In addition to the spectra which appear in the orthorhombic lattice, 
new spectra marked by $M$ are found. 
Spectral intensity of the peaks $M$ are small compared with others.
This is because the distortion from the orthorhombic structure is small. 
The new peaks at 3.6~$J_1$ and 4.4~$J_1$ originate from the fact that the mirror symmetry 
perpendicular to the $z$ axis is absent in the monoclinic structure $P2_1/c$. \cite{mitchell} 
Most strikingly, there appears a new peak at 3.1~$J_1$. 
The peak position corresponds to the energy of the orbital wave at the X point 
marked by an asterisk in the Fig.~\ref{fig:fig2} (b). 
This mode becomes zone center mode in the Brillouin zone for the monoclinic lattice 
and becomes Raman-active. 
\par
Now, let us compare our theoretical results with the experimental ones. 
Recently, Saitoh {\it et al.} have reported the experimental results of the Raman scattering 
in a detwinned single crystal of LaMnO$_3$. \cite{e.saitoh}
In addition to the phonon Raman spectra below 100~meV,\cite{yamamoto,iliev}
sharp spectra in several polarization configurations at 120$\sim$170~meV are observed 
as shown in the insets of Fig.~\ref{fig:fig7}. 
The Raman spectra observed in such a high energy region are usually attributed to 
the multi-phonon excitations in transition-metal oxides.\cite{romero} 
However, by comparing the polarization and the temperature dependence 
of the new spectra with those of the phonon ones, 
the possibility of the multi-phonon excitation is ruled out. 
Two-magnon excitations are also ruled out because 
spins are ferromagnetically aligned in the $xy$ plane below $T_N$. 
As a result, orbital wave is considered as a candidate for the remaining excitation. 
The crystal structure of the sample in these experiments is monoclinic 
because of the oxygen partial pressure during synthesis.\cite{mitchell} 
In this experiments, the 514.5~nm line (2.4~eV) of an Ar$^+$ laser was used. 
This energy 2.4~eV corresponds to the charge-transfer excitation 
from O~$2p$ to Mn~$e_g$ orbitals.\cite{arima} 
Therefore, these experimental results are compared with our theoretical ones 
in the $d$-$p$ process in the monoclinic lattice. 
As shown in Fig.~\ref{fig:fig7}, 
the characteristic features of polarization dependence and relative intensity 
of the experimental Raman spectra in the region of 120$\sim$170~meV 
are well reproduced by our theoretical ones. 
Thus, the Raman spectra observed in the region of 120$\sim$170~meV in LaMnO$_3$ 
are attributed to the orbital wave excitations. 
%
%
%
%
%
%
%
%
%
\section{Summary and Discussion}
In this study, we have theoretically investigated the Raman scattering 
as a probe to detect the orbital wave excitations in orbital ordered manganites. 
Two excitation processes were proposed for the Raman scattering, 
i.e., the $d$-$d$ and $d$-$p$ processes.  
The $d$-$d$ process is analogous to the two-magnon Raman scattering process in the antiferromagnet. 
However, scattering intensity from one- and two-orbital wave excitations are 
of the same order of magnitude unlike the two-magnon Raman scattering. 
This is due to the transfer intensity between the NN different orbitals. 
In the $d$-$p$ process, 
photon induces an exchange of electrons between Mn $e_g$ and O $2p$ orbitals,  
and one-orbital excitations are brought about. 
Because LaMnO$_3$ is a charge-transfer type insulator where the optical gap is about 1~eV, 
the $d$-$p$ process is expected to dominate the Raman scattering when the visible light is used. 
It was shown that the theoretical results of the Raman spectra from the one-orbital wave excitations 
well explain the experimental spectra observed in the region of 120$\sim$170~meV in LaMnO$_3$. 
\par
As mentioned in Sec.~II, the orbital wave in the $A$-AF phase has a gap originating from 
the anisotropic spin structure. 
Therefore, we expect that the gap is suppressed by applying a magnetic field. 
This change will be reflected on the peak positions of the orbital-wave Raman spectra and 
will be experimentally detected. 
Similarly, the gap-less orbital wave excitation may be observed in 
the ferromagnetic-insulating manganites such as La$_{0.88}$Sr$_{0.12}$MnO$_3$ 
where the orbital ordering is experimentally confirmed.\cite{endoh}
In Sec.~V, we compared the theoretical results for the $d$-$p$ process 
with the recent experiments in Ref.~\onlinecite{e.saitoh}. 
This is because the $d$-$p$ process is dominant in the adopted energy of the incident photon. 
By changing the incident photon energy, the $d$-$d$ process may contribute to the scattering. 
It is expected that 
the spectral intensities for the $d$-$d$ process are enhanced 
with increasing the energy of the incident photon and 
resonating with the excitation energy from the occupied lower Hubbard band to the unoccupied 
upper Hubbard one. 
When the Raman experiment using such a high-energy photon is carried out, 
we expect that the comparison between the present theoretical results for the $d$-$d$ process 
and experiments will provide much information about the orbital wave. 
It is expected that the Raman spectra from the two-orbital wave excitation will be observed 
around 200$\sim$350~meV in LaMnO$_3$. 
\acknowledgments
The authors would like to thank Y.~Tokura, E.~Saitoh, T.~Takahashi, K.~Tobe, K.~Yamamoto and T.~Kimura 
for their valuable discussions and providing us experimental data before publication. 
Fruitful discussions with P.~Prelov{\v s}ek, G.~Khaliullin and K.~Tsuda are also acknowledged. 
This work was supported by Grant-in-Aid for Scientific Research Priority Area from 
the Ministry of Education, Science, Sports, Culture and Technology of Japan, CREST Japan 
and Science and Technology Special Coordination Fund for Promoting Science and Technology. 
Part of the numerical calculation was performed in 
the supercomputing facilities in IMR, Tohoku University. 
One of the authors (S. M.) acknowledges support of the Humboldt Foundation. 


\begin{references}
\bibitem{JJ}
$J_1$ and $J_2$ are defined by 
$J_1 = t_0^2/(U' - I)$ and $J_2 = t_0^2/(U' + I + 2J_H)$, respectively, 
where a relation $U = U' + I$ is assumed. 
$t_0$ is the transfer intensity between NN $d_{3z^2-r^2}$ orbitals along the $z$ axis. 
Energy splitting due to the JT effect is neglected in the denominators 
because the splitting is much smaller than the Coulomb interactions.
\end{references}
\end{document}